\documentclass[aps,prd,twocolumn,showpacs,superscriptaddress,preprintnumbers]{revtex4-1} 
\usepackage[hypertex,colorlinks=true, pdfstartview=FitV, linkcolor=red,citecolor=blue, urlcolor=blue,
bookmarks=true, bookmarksnumbered=true, dvipdfmx]{hyperref}
\usepackage[dvipdfmx]{graphicx}
\usepackage{amsmath,amssymb}
\usepackage{tabularx}
\usepackage{ulem}
\usepackage{mathrsfs}
\usepackage{subfigure}
\usepackage{braket}
\def\vec#1{\mbox{\boldmath $#1$}}

\begin{document}

\title{The radiative capture reaction rate from $\Lambda \Lambda $ to H dibaryon\\ in the imaginary time method}

\author{Eri Hikota}
\email{hikoeri@th.phys.titech.ac.jp}
\affiliation{Department of Physics, H-27, Tokyo Institute of Technology, Meguro, Tokyo 152-8551, Japan}
\author{Yasuro Funaki}
\email{funaki@riken.jp}
\affiliation{RIKEN Nishina Center, Hirosawa 2-1, Wako, Saitama 351-0198, Japan}
\author{Emiko Hiyama}
\email{hiyama@riken.jp}
\affiliation{RIKEN Nishina Center, Hirosawa 2-1, Wako, Saitama 351-0198, Japan}
\author{Makoto Oka}
\email{oka@th.phys.titech.ac.jp}
\affiliation{Department of Physics, H-27, Tokyo Institute of Technology, Meguro, Tokyo 152-8551, Japan}
\affiliation{Advanced Science Research Center, JAEA, Tokai, Ibaraki 319-1195, Japan}

\begin{abstract}
Radiative capture rates of thermal $\Lambda\Lambda + \Xi$N states into H dibaryon are calculated in the novel imaginary time method.
The H dibaryon is assumed to be a bound state of $\Xi $N with spin $J^{\pi}= 0^+$, isospin $I=0$ and strangeness $-2$.
We consider $E1$ transition to H from $\Xi$N $(L=1)$ scattering states which mix with $\Lambda\Lambda (L=1)$.
In order to calculate the transition rates, we formulate a coupled-channel imaginary time method by extending the one-channel
formula originally proposed by Yabana and Funaki. 
The imaginary time method allows us to avoid the
sum over all the excited thermal initial states, and thus to save computational time significantly.
The transition rates are given as a function of temperature and the unknown 
binding energy of the H dibaryon, which we take as a parameter. 
It is found that the transition rate is not sensitive to the choices of the H binding energy or the strengths of the channel coupling 
for temperatures 3 MeV or higher.
\end{abstract}
\maketitle

\section{Introduction}
H dibaryon is a compact bound (or resonance) state of six quarks, $u^2d^2s^2$, with spin$=0$, isospin$=0$ and strangeness$=-2$, proposed by Jaffe in 1977\cite{Jaffe}.
The lowest two baryon threshold with these quantum numbers is the S-wave, spin-singlet $\Lambda\Lambda$ ($L=0, J=0$) state. 
The reason why the H dibaryon is interesting is that the quark model dynamics, which describes the ground state hadrons very well, suggests
a strong attractive force in this channel. 
In particular, the flavor SU(3) singlet configuration of the six quarks is favored most by the color-magnetic force ($\propto -(\vec{\sigma}_i\cdot\vec{\sigma}_j)(\lambda_i\cdot\lambda_j)$), 
which is responsible for the spin splittings of the pseudo-scalar and vector mesons as well as the octet $1/2^+$ and the decuplet $3/2^+$ baryons. 
It is also shown that the Fermi-Dirac statistics for quarks allows all six quarks can occupy the lowest orbit making a fully symmetric orbital state and therefore no Pauli blocking effect may induce short-distance repulsion among the quarks.
Thus a compact six-quark-like state is expected in the quark models\cite{Oka:1983ku,Takeuchi:1991,Sakai:1999qm}. 

The quantum chromodynamics (QCD) also allows color-singlet six-quark states.
Recent lattice calculations have indeed suggested existence of either a bound or resonance state around the $\Lambda\Lambda$ threshold, although the physical quark-mass point calculations are yet to come\cite{NPLQCD,HALQCD}.
Against all these theoretical indications, long experimental efforts in searching the H dibaryon were unsuccessful so far. The observation of the double hypernucleus has given an upper bound for the binding energy\cite{Takahashi}, while an enhancement is observed in the $\Lambda\Lambda$ production around and above the threshold.\cite{Yoon}. 

Under these circumstances, it is interesting to explore possibilities of finding the H dibaryon in various production mechanism\cite{EXHIC}. 
In this paper, we investigate the radiative fusion production of H, $\Lambda+\Lambda \to \gamma \textnormal{H}$, in an environment of finite temperature baryon gas. Suppose that a heavy ion collision produces significant number of strange baryons in a temperature $T$. Then through the radiative fusion, the H dibaryons may be produced associated with emitted characteristic photons.
The goal is to calculate the production rate from the finite-$T$ $\Lambda$ gas into H dibaryons.
In the present calculation we assume that our final state is a $\Xi$N bound state and the radiative transition is through mixing of $\Xi$N (intermediate) states with $\Lambda\Lambda$ states.

Our setup here is as follows. The final state is a spin-singlet S-wave bound state of $\Xi$ and N. The radiative decay occurs most strongly from the P-wave $\Xi$N ($J=1$) states. We consider both the spin singlet (${}^1\textnormal{P}_1$) and the spin-triplet (${}^3\textnormal{P}_1$) $\Xi$N states. They mix with the P-wave $\Lambda\Lambda$ states, which due to the Pauli principle is always spin-triplet (${}^3\textnormal{P}_1$). 
Then overall, the transition goes from $\Lambda\Lambda({}^3\textnormal{P}_1)$ to $\Xi$N$ ({}^1\textnormal{P}_1, {}^3\textnormal{P}_1)$ to $\gamma + \textnormal{H}(\Xi \textnormal{N}, {}^1\textnormal{S}_0)$.

In order to calculate the production rate, we apply the imaginary time method (ITM)~\cite{Yabana}, in which imaginary time is identified with inverse temperature. This method allows us to replace the sum of the initial scattering-state distribution at finite temperature by solving a Schr\"odingier-type differential equation with the imaginary-time variable. The advantage of the ITM is that it takes into account the contributions of all the excited initial states automatically without approximation and explicit treatment of them, for instance, discretizing the continuum energy spectrum.

This advantage was demonstrated in the study of the triple-alpha radiative capture process~\cite{Ya14}, which plays a key role in synthesizing ${^{12}{\rm C}}$ in stars~\cite{Sa52,Ho54}. 
In particular, at low temperatures the reaction rate of this process is very difficult to calculate, since direct capture of the three-alpha particles through their scattering states becomes important, but formal scattering theory of the three-charged particles is not available. This is in contrast with the situation at high temperatures, in which the reaction proceeds through the so-called Hoyle state, the resonant $0^+$ state at $7.65$ MeV in ${^{12}{\rm C}}$~\cite{Co57,Ho54}. However, the ITM that does not require any boundary condition of the three-charged particles could overcome this difficulty and was found to be very powerful method.

In this study, we extend the ITM mentioned above in coupled-channel calculations, in which the three-channel initial states, $\Lambda\Lambda(^3\textnormal{P}_1)$ and $\Xi $N$({}^1\textnormal{P}_1, {}^3\textnormal{P}_1)$, are considered. We show that the coupled-channel ITM reduces conputational time significantly, since it does not require solving scattering initial states explicitly. The calculated transition rate at high temperatures is shown to be insensitive to the energy positions of the H dibaryon, though the H dibaryon is assumed to be bound in this calculation, due to a constraint of the present method.



This paper is organized as follows.

In sect. 2, we summarize the formulation of the coupled-channel ITM.
The hamiltonian for the initial states is taken from the Nijmegen potentials supplemented by an anti-symmetric spin-orbit (ALS) force.  The ALS is necessary to mix the spin 0 $\Xi$N state with spin 1.
The final state energy, i.e., the H dibaryon mass, and its wave function are computed by assuming a simple Gaussian potential for ${}^1\textnormal{S}_0$ $\Xi$N system.


In sect. 3, we present details of our calculation, solving the imaginary time differential equation step by step. Our main results are shown, i.e., the reaction rates for various temperatures, contributions of the initial channels, as well as dependencies of the results on the choices of parameters of the model.

In sect. 4, conclusions are given.
%
%
%
%
\section{Imaginary time method}
In this paper, we consider the transition process in the heavy ion collisions.
As mentioned in Introduction, H dibaryon is assumed to be $^1\textnormal{S}_0$ bound state of $\Xi $ and N as the final state of the radiative capture reaction. $\Lambda \Lambda$ states are thus not assumed to change into the H dibaryon directly by emitting a photon, though they are thermally distributed as the initial states. The initial states that are most likely to decay into the final state are then $\Xi $N(${}^3\textnormal{P}_1$ and $^1\textnormal{P}_1$) states, which mix with the $\Lambda \Lambda$ $({}^3\textnormal{P}_1)$ states by the strong interaction. This transition may be uniquely indentified as a photon with a few tens of MeV is emitted.

The ordinary form of the radiative capture thermal reaction rate accompanying an emission of a photon between the initial and final states, which are labeled by $i,c$ and $f$, respectively, is given by
\begin{eqnarray}
r \left( \beta \right)&=& \frac{1}{\omega _i} \left( \frac{2\pi \beta \hbar ^2}{\mu }\right) ^{3/2} \sum_{M_f\mu } \sum_i e^{-\beta E_i} \frac{8\pi \left( \lambda +1\right) }{\hbar \lambda \left[ \left( 2\lambda +1\right) !!\right] ^2}\nonumber \\
&&\hspace{1cm}\times \left( \frac{E_{\gamma }}{\hbar c}\right) ^{2\lambda +1} \left| \sum_{c}\Braket{\psi _f|\hat{M}_{\lambda \mu }|\psi _{ic}}\right| ^2. \nonumber \\ \label{ordinary}
\end{eqnarray}
Here $\beta =1/k_{\textnormal{B}}T$ is the inverse temperature, $E_{\gamma }=E_i-E_f$ is the energy of the emitted photon, and $\psi _{ic}$ and $\psi _f$ are the wave functions of the initial and final states, respectively, in which $c$ specifies the channel number of the initial states, i.e. $c=1,2$ and $3$ for $\Lambda \Lambda (^3\textnormal{P}_1)$, $\Xi $N$(^1\textnormal{P}_1)$ and $\Xi $N$(^3\textnormal{P}_1)$ states, respectively. $\hat{M}_{\lambda \mu }$ is the multipole transition operator for $\gamma$-ray emission with a multipolarity $\lambda$. Since we now consider the reaction with total angular momentum $J=1$, a $\gamma$-ray with $\lambda=1$ is emitted, giving the transition to the ${}^1S_0$ H dibaryon. $\omega _i$ accounts for the degeneracy of the initial state, i.e. $\omega_i=2\cdot 1 +1$, and $\mu $ is the reduced mass of the initial state, which is taken as the reduced mass of $\Xi \textnormal{N}$.

In order to obtain a form of the ITM, it is convenient to use the following spectral representation of the Hamiltonian,
\begin{eqnarray}
&&\left\{  e^{-\beta \hat{H}} \left( \frac{\hat{H}-E_f}{\hbar c}\right) ^{3}\right\} _{cc'} \nonumber \\
&=&\sum_{n\in \textnormal{bound}}e^{-\beta E_n}\left( \frac{E_n-E_f}{\hbar c}\right)^{3}\Ket{\psi _{nc}} \Bra{\psi _{nc'}} \nonumber \\
&&+\sum_{i\in \textnormal{scattering}}e^{-\beta E_i}\left( \frac{E_i-E_f}{\hbar c}\right)^{3}\Ket{\psi _{ic}} \Bra{\psi _{ic'}}.
\end{eqnarray}
By substituting the above equation into the ordinary form of the reaction rate Eq.~(\ref{ordinary}), the following ITM formula can be obtained,
\begin{eqnarray}
r&&\left( \beta \right) =\frac{16\pi }{9\hbar } \left( \frac{2\pi \beta \hbar ^2}{\mu _{\Xi \textnormal{N}}} \right) ^{3/2} \sum_{\mu } \nonumber \\
&&\times \Braket{\psi _f|\hat{M}_{\lambda \mu } e^{-\beta \hat{H}}\left( \frac{\hat{H}-E_f}{\hbar c}\right) ^{3} \hat{P} \hat{M}_{\lambda \mu}^\dagger |\psi _f}, \label{eq:itm} \nonumber \\
\end{eqnarray}
where
$\hat{H}$ is the Hamiltonian for the three-component initial states and $\hat{P}$ is the projection operator that eliminates any bound initial states. In the present calculations, the Hamiltonian does not give any bound eigenstates, which leads to $\hat{P}=1$. 

The transition operator that we adopt in the present calculations is given by the standard (leading-order) $E1$ ($\lambda=1$) transition operator and a sub-leading operator, which flips the spin of the particle, {\it i.e.},
\begin{equation}
\hat{M}_{\lambda \mu}=\int \rho r^\lambda Y_{\lambda \mu}^*(\hat{\vec{r}}) d\vec{r} -\frac{ik}{\lambda+1}\int \nabla \cdot (\vec{r}\times \vec{m})r^\lambda Y_{\lambda \mu}^* (\hat{\vec{r}})d\vec{r},
\end{equation}
where $k$ is the momentum of the emitted photon, $\rho$ represents the charge density, and $\vec{m}$ denotes the magnetic moment density.
The matrix elements of this operator for the individual initial channels are given by
\begin{eqnarray}
\hspace{-0.4cm} \langle \psi_f| \hat{M}_{\lambda=1 \mu } | \psi_{ic=1} \rangle &=& 0,\label{eq:tra1} \\
\hspace{-0.4cm} \langle \psi_f| \hat{M}_{\lambda=1 \mu } | \psi_{ic=2} \rangle &=&-\frac{e}{4\sqrt{\pi }}\int_0^\infty dr\ r R_f(r)R_{ic=2}(r), \label{eq:tra2} \\
\hspace{-0.4cm} \langle \psi_f| \hat{M}_{\lambda=1 \mu } | \psi_{ic=3} \rangle &=&\frac{gk}{2\sqrt{2\pi }}\int_0^\infty dr\ r R_f(r)R_{ic=3}(r), \label{eq:tra3}
\end{eqnarray}
where $e$ is the electric charge, and $R_{ic}(r)$ and $R_f(r)$ are radial wave functions of the initial and final states, respectively.
Note that the leading-order transition operator cannot change the spin, while the sub-leading term connects the spin 1 state to the spin 0 final state.
We here assume no exchange current operator which may change $\Lambda\Lambda$ to $\Xi$N in the transition.
The magnetic coupling $g$ can be expressed in terms of the iso-scalar part of the $g$-factors of $\Xi$ and N, $g_1$ and $g_2$, respectively, in the following way,
\begin{eqnarray}
g=\frac{g_1\frac{m_2}{m_1+m_2}-g_2\frac{m_1}{m_1+m_2}}{2}=-0.1572\mu _{\textnormal{N}}.
\end{eqnarray}
Here $m_1$ and $m_2$ are masses of $\Xi$ and N particles, respectively, and $\mu _{\textnormal{N}}$ is the nuclear magneton. 

The Hamiltonian $\hat{H}$ for the initial states can be described as having the three-channel components. The potential parts, which we denote as $v_{cc'}$ with $c,c'=1,2,3$, can then be composed of the central, spin-orbit (SLS), antisymmetric spin-orbit (ALS), and tensor terms, as follows:
\begin{eqnarray}
v_{cc'}&=&\sum_{k}v_k^{\textnormal{(C)}} e^{-\left( r/r_{k}^{\textnormal{(C)}}\right) ^2}
+\mbox{\boldmath $L$}\cdot \mbox{\boldmath $S$}^{(S)}\sum_{k}v_k^{\textnormal{(S)}} e^{-\left( r/r_{k}^{\textnormal{(S)}}\right) ^2} \nonumber \\
&+&\mbox{\boldmath $L$}\cdot \mbox{\boldmath $S$}^{(A)}\sum_{k}v_k^{\textnormal{(A)}} e^{-\left( r/r_{k}^{\textnormal{(A)}}\right) ^2}
+S_{\textnormal{T}}\sum_{k}v_k^{\textnormal{(T)}} e^{-\left( r/r_{k}^{\textnormal{(T)}}\right) ^2}, \nonumber \\ \label{gauss}
\end{eqnarray}
where the SLS, ALS and tensor operators, $\vec{L}\cdot \vec{S}^{(S)}$, $\vec{L}\cdot \vec{S}^{(A)}$ and $S_T$, respectively, can be defined as,
\begin{eqnarray}
\mbox{\boldmath $L$}\cdot \mbox{\boldmath $S$}^{(S)}&=&
\mbox{\boldmath $L$}\cdot \left( \mbox{\boldmath $S$}_1+\mbox{\boldmath $S$}_2\right)
=\mbox{\boldmath $L$}\cdot \left( \frac{\mbox{\boldmath $\sigma $}_1+\mbox{\boldmath $\sigma $}_2}{2}\right), \\
\mbox{\boldmath $L$}\cdot \mbox{\boldmath $S$}^{(A)}&=&
\mbox{\boldmath $L$}\cdot \left( \mbox{\boldmath $S$}_1-\mbox{\boldmath $S$}_2\right)
=\mbox{\boldmath $L$}\cdot \left( \frac{\mbox{\boldmath $\sigma $}_1-\mbox{\boldmath $\sigma $}_2}{2}\right), \\
S_{\textnormal{T}}&=&S_{12}=3\left( \mbox{\boldmath $\sigma $}_1\cdot \hat{r}\right) \left( \mbox{\boldmath $\sigma $}_2\cdot \hat{r}\right)
-\mbox{\boldmath $\sigma $}_1\cdot \mbox{\boldmath $\sigma $}_2.
\end{eqnarray}
Here we note that for the YN (hyperon-nucleon) and YY (hyperon-hyperon) interaction in Eq.~(\ref{gauss}) we adopt $G$-matrix-type YN and YY interaction constructed from Nijmegen potential. For the $G$-matrix-type interaction, which is described in terms of multi-range Gaussian form, we take the parameter set of NSC-97f~\cite{hiyamasan_ronbun} for the ALS term and those of ESC08c~\cite{Ya15} for the other terms. The detailed list of the parameter sets is shown in Appendix.

It should be noted that the ALS term does not appear in $v_{11}$, $v_{22}$, $v_{33}$ and $v_{13}=v_{31}$ and only contributes to $v_{12}=v_{21}$ and $v_{23}=v_{32}$, and no terms appear in $v_{12}=v_{21}$ and $v_{23}=v_{32}$ other than the ALS term.
Furthermore, one can prove that under SU(3) symmetry limit the following relation is satisfied, 
\begin{eqnarray}
v_{\textnormal{ALS}}=v_{12}=v_{21}=v_{23}=v_{32}=-2v_{\Lambda \textnormal{N}}, \label{eq:vals}
\end{eqnarray}
where $v_{\Lambda \textnormal{N}}$ is the ALS potential between $\Lambda \textnormal{N}(^3\textnormal{P}_1)$ and $\Lambda \textnormal{N}(^1\textnormal{P}_1)$ states.

\begin{figure}[htbp]
\begin{center}
\includegraphics[width=7.5cm, height=5.5cm]{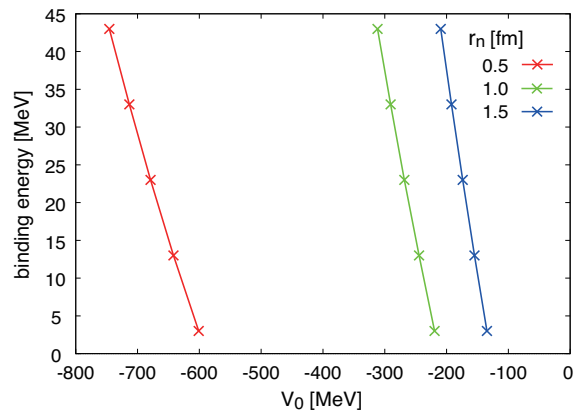}
\caption{Binding energies for various choices of potential parameters of the H dibaryon, $r_{\textnormal{H}}$ and $V_0$ in Eq.~(\ref{eq:h}).}
\label{Hdibaryon_pot_para}
\end{center}
\end{figure}
\begin{figure}[htbp]
\begin{center}
\includegraphics[width=7.5cm, height=5.5cm]{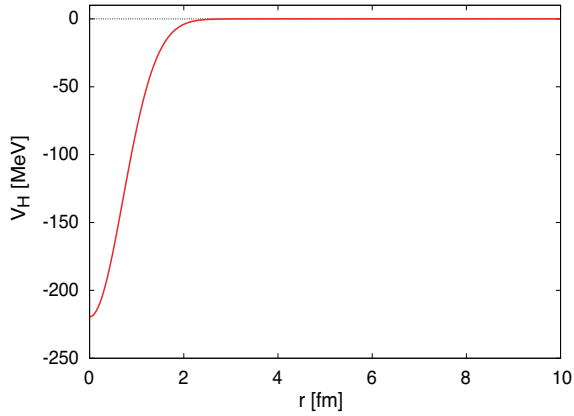}
\caption{The potential of the H dibaryon with $V_0=-219.44$ [MeV] and $r_{\textnormal{H}}=1.0$ [fm]. The horizontal axis represents a relative distance between the $\Xi$ and N particles.}
\label{pot_Hdibaryon}
\end{center}
\end{figure}
\begin{figure}[htbp]
\begin{center}
\includegraphics[width=7.5cm, height=5.5cm]{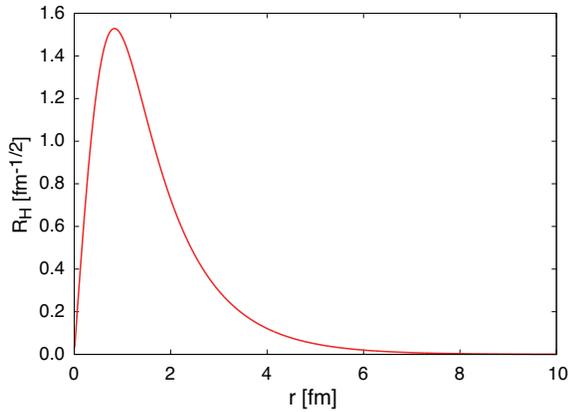}
\caption{The radial part of the wave function of the H dibaryon. The horizontal axis represents a relative distance between the $\Xi$ and N particles.}
\label{Hdibaryon_wf}
\end{center}
\end{figure}

Next we calculate the final-state wave function by solving the Schr\"odinger equation, assuming that the H dibaryon is a $\Xi $N bound state.
Here we simply adopt a Gaussian form for the potential of the H dibaryon as follows:
\begin{eqnarray}
V_{\textnormal{H}}=V_0e^{-\left( r/r_{\textnormal{H}}\right) ^2}, \label{eq:h}
\end{eqnarray}
where $r_{\textnormal{H}}$ is a width parameter corresponding to the size of the H dibaryon, for which we take $r_{\textnormal{H}}=0.5, 1.0, 1.5$\ [fm].
The size of the width parameter $r_{\textnormal{H}}=0.5$\ [fm] corresponds to the compact 6 quarks H dibaryon and $r_{\textnormal{H}}=1.5$\ [fm] corresponds to a $\Xi $N bound state with more dilute density structure.
Figure~\ref{Hdibaryon_pot_para} shows the relations between the various choices of $r_{\textnormal{H}}$ and $V_0$, and the binding energies of the H dibaryon from the $\Lambda \Lambda $ threshold.
In the subsequent calculations, we adopt $r_{\textnormal{H}}=1.0$ [fm] and $V_0=-219.44$ [MeV], giving the binding energy of the H dibaryon $E_f=3.0$ [MeV] from the $\Lambda \Lambda$ threshold. The potential curve for this choice of the parameters  $r_{\textnormal{H}}$ and $V_0$ is shown in Fig.~\ref{pot_Hdibaryon}. 
A radial grid size of $\Delta r=0.01$ [fm] is taken in all the calculations in this work, for which we confirmed that the present results sufficiently converge. The radial wave function of the H dibaryon with this potential is also shown in Fig. \ref{Hdibaryon_wf}.

In order to calculate the reaction rate via the ITM in Eq.~(\ref{eq:itm}), it is useful to first define the following wave function with three components $\vec{\Psi}(\beta)=(\Psi_1(\beta),\Psi_2(\beta),\Psi_3(\beta))$ depending on the inverse temperature $\beta$ that is identified with the imaginary-time variable,
\begin{eqnarray}
\vec{\Psi} \left(\beta \right) =e^{-\beta \hat{H} }\hat{M}^{\dagger }_{\lambda =1 \mu } \psi _f.
\label{eq:itm_wf}
\end{eqnarray}
This wave function is found to satisfy the following Schr\"odinger-type equation along the imaginary-time axis $\beta$,
\begin{eqnarray}
-\frac{\partial }{\partial \beta }\vec{\Psi} \left( \beta \right) =\hat{H} \vec{\Psi} \left( \beta \right). \label{imaginary_eq}
\end{eqnarray}
The formula in Eq.~(\ref{eq:itm}) can then be expressed by using the imaginary-time wave function $\vec{\Psi}(\beta)$ as follows:
\begin{eqnarray}
r&&\left( \beta \right) =\frac{16\pi }{9\hbar } \left( \frac{2\pi \beta \hbar ^2}{\mu _{\Xi \textnormal{N}}} \right)^{3/2} \sum_{\mu }\sum_{cc'} \nonumber \\
&&\times \Braket{\Psi_c(\beta/2)| \left( \frac{\hat{H}-E_f}{\hbar c}\right)^{3}_{cc'} |\Psi_{c'}(\beta/2)}. \label{eq:itm2}
\end{eqnarray}
The reaction rate at an arbitrary value of $\beta$ can be obtained as a result of the imaginary-time evolution of $\vec{\Psi}(\beta)$ by solving Eq.~(\ref{imaginary_eq}) with the use of the Taylor expansion method, in the following way,
\begin{eqnarray}
\vec{\Psi} \left( \beta+\Delta \beta \right) &=&e^{-\Delta \beta \hat{H}}\vec{\Psi} \left( \beta \right) \nonumber \\
&\simeq &\sum_{k=0}^{k_{\rm max}}\frac{\left( -\Delta \beta \hat{H} \right) ^k}{k!} \vec{\Psi} \left(\beta \right), \label{taylor}
\end{eqnarray}
where we take $\Delta \beta =1\times 10^{-6}$ [MeV$^{-1}$] and $k_{\rm max}=4$. We confirmed that with this choice of the variables the results well converge. 


\section{Results and discussion}

\begin{figure}[htbp]
\begin{center}
\subfigure[The $\Lambda \Lambda$ component $\Psi_{c=1}(\beta =0)$.]{%
\includegraphics[width=7.5cm, height=5.5cm]{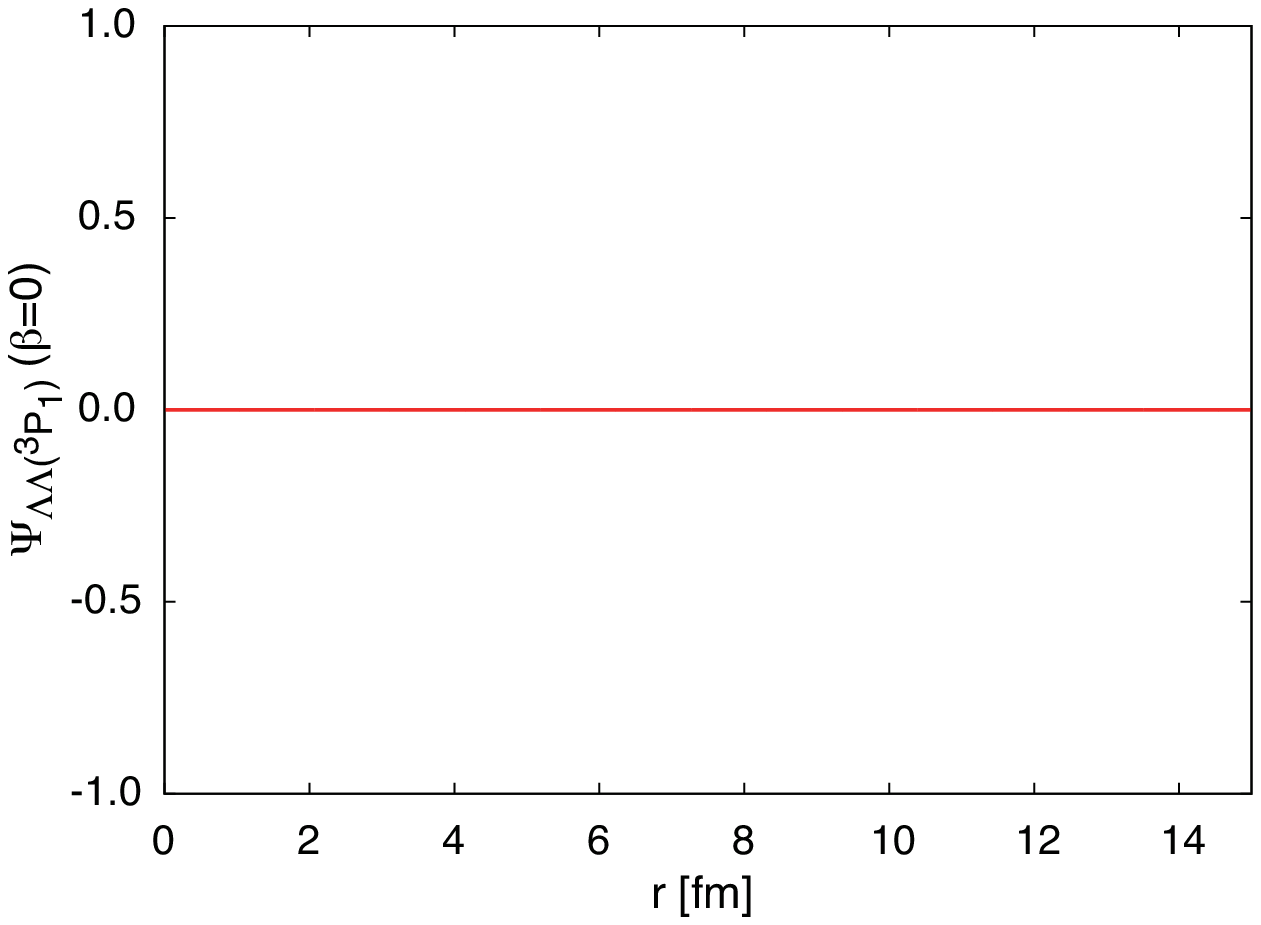}}%
\hspace{10mm}
\subfigure[The $\Xi $N$(^1\textnormal{P}_1)$ component $\Psi_{c=2}(\beta =0)$.]{%
\includegraphics[width=7.5cm, height=5.5cm]{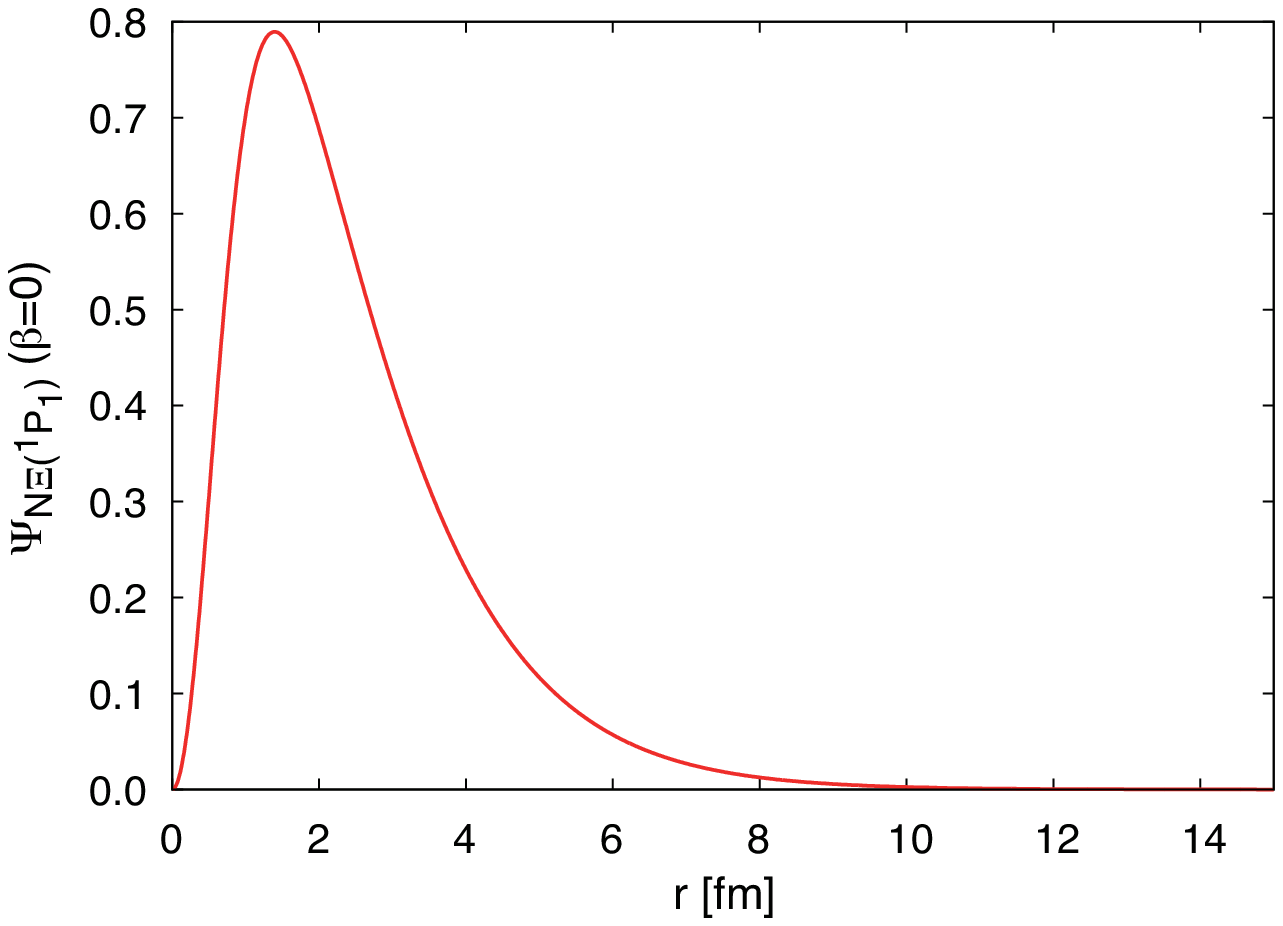}}%
\hspace{10mm}
\subfigure[The $\Xi $N$(^3\textnormal{P}_1)$ component of $\Psi_{c=3}(\beta =0)$.]{%
\includegraphics[width=7.5cm, height=5.5cm]{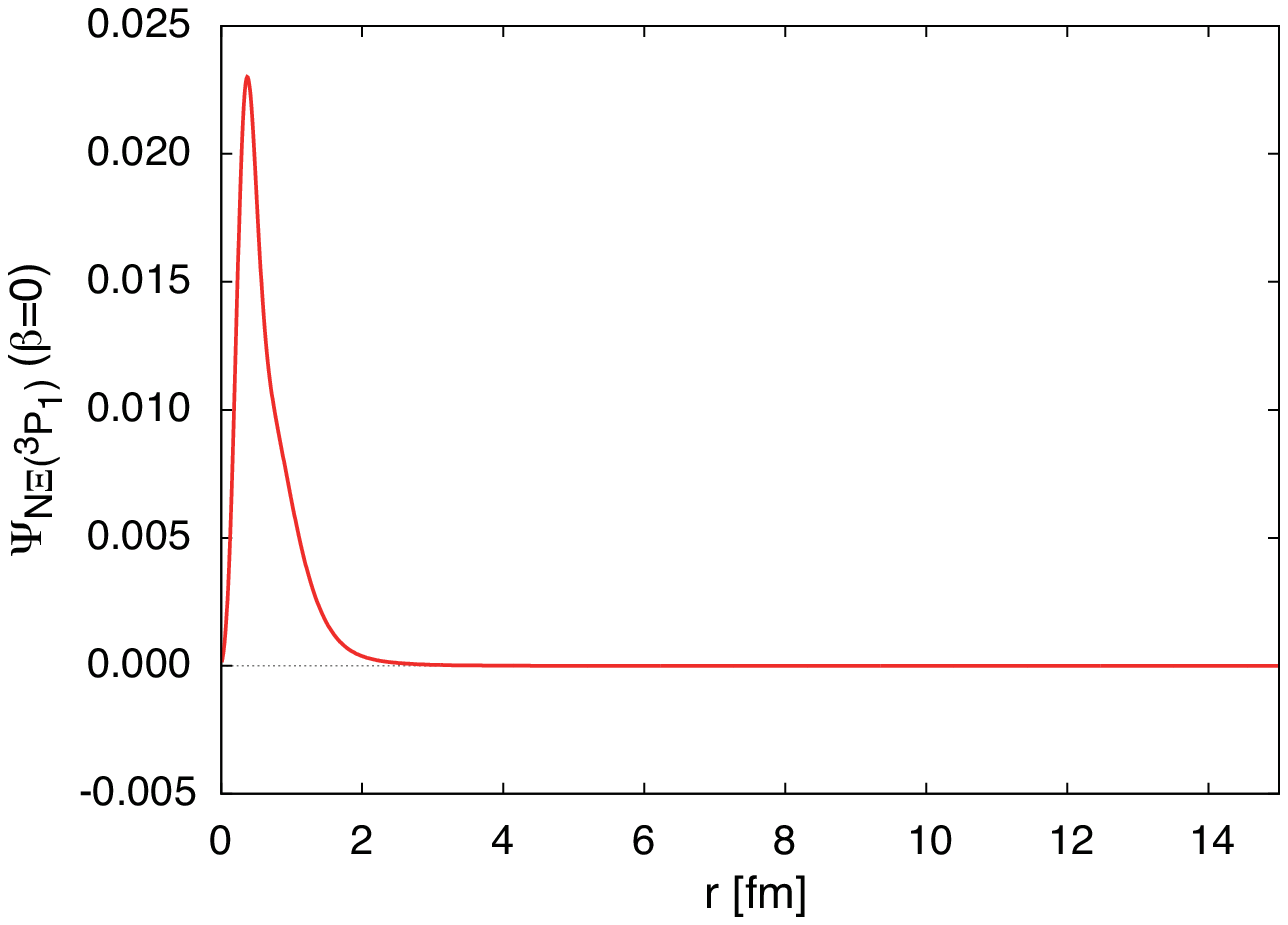}}%
\caption{The radial part of the three-component wave function $\vec{\Psi} (\beta)$ in Eq.~(\ref{eq:itm_wf}) at $\beta=0$.}
\label{psi_beta0}
\end{center}
\end{figure}

Figure~\ref{psi_beta0} shows the initial distributions with $\beta=0$, of the $\Lambda \Lambda $, $\Xi $N($^1\textnormal{P}_1$), and $\Xi $N($^3\textnormal{P}_1$) components of the $\vec{\Psi} (\beta /2=0)$ wave function in Eq.~(\ref{eq:itm_wf}), $\Psi_1(\beta/2=0)$, $\Psi_2(\beta/2=0)$ and $\Psi_3(\beta/2=0)$, respectively.
We can see that there is no $\Lambda \Lambda $ component, i.e. $\Psi_1(\beta /2=0)=0$, since the transition operator $\hat{M}_{\lambda=1 \mu}$ does not couple the final state with the $\Lambda \Lambda $ initial state, as calculated in Eq.~(\ref{eq:tra1}). The amplitudes of $\Xi $N($^1\textnormal{P}_1$) and $\Xi $N($^3\textnormal{P}_1$) components are non-zero, and the former is larger than the latter, together with a broader shape of the former component, since the factor $k$ in the transition operator for the former channel in Eq.~(\ref{eq:tra2}) enhances the amplitude of the wave function at higher temperature region.
%

\begin{figure}[htbp]
\begin{center}
\subfigure[The wave function at $\beta=0.0001$ MeV$^{-1}$.]{%
\includegraphics[width=7.5cm, height=5.5cm]{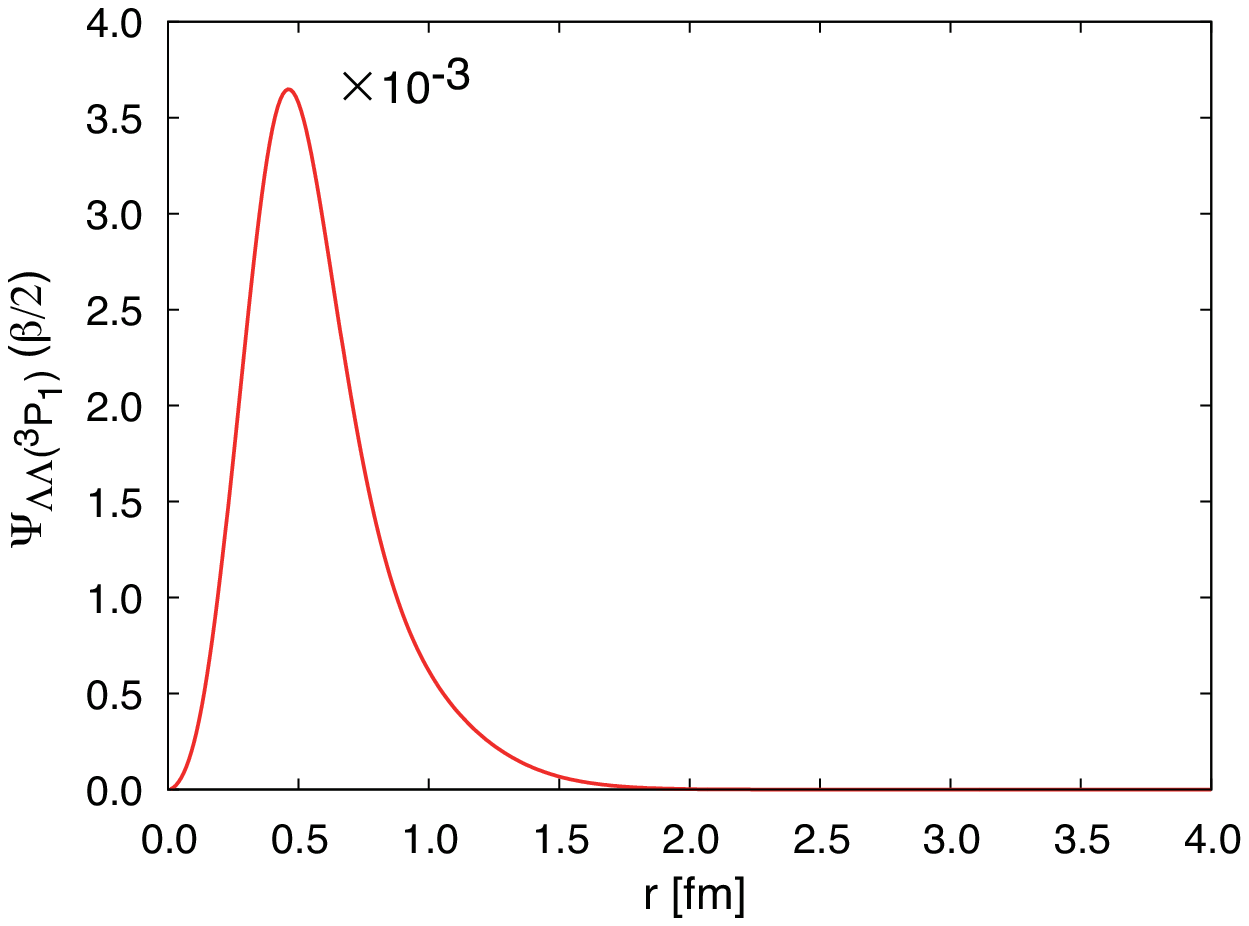}}%
\hspace{10mm}
\subfigure[The wave function at $\beta =0.001$ MeV$^{-1}$.]{%
\includegraphics[width=7.5cm, height=5.5cm]{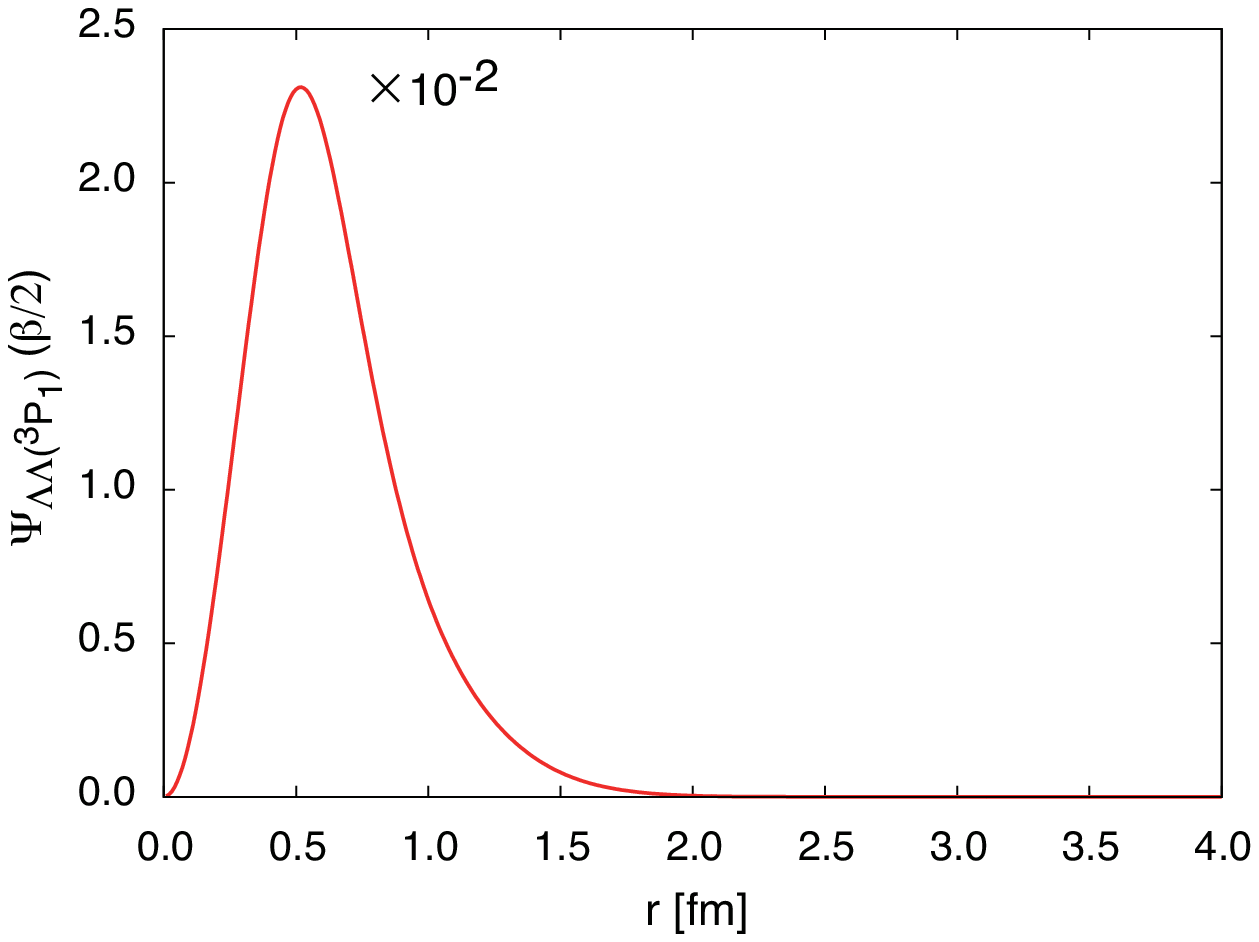}}%
\hspace{10mm}
\subfigure[The wave function at $\beta =0.01$ MeV$^{-1}$.]{%
\includegraphics[width=7.5cm, height=5.5cm]{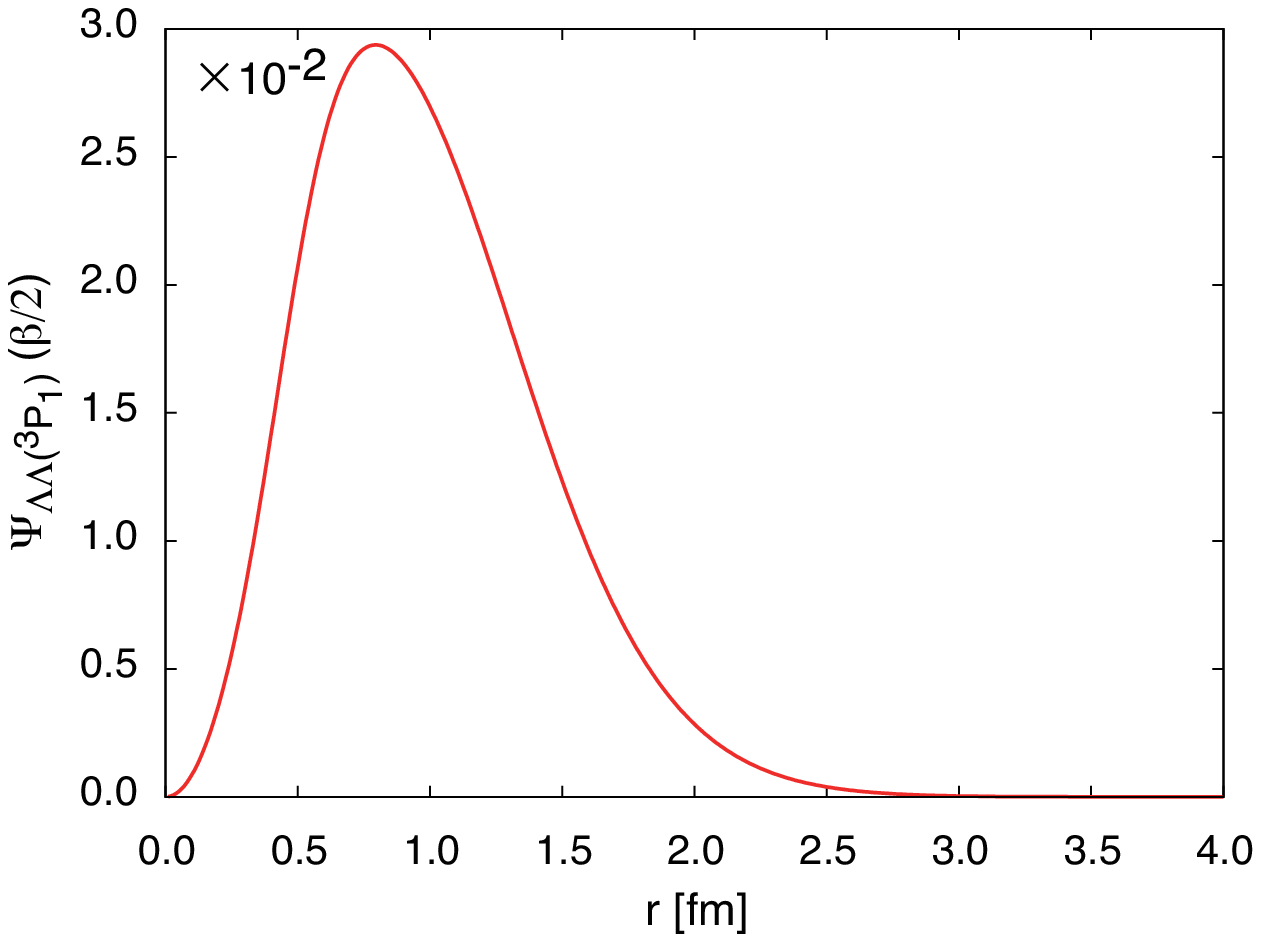}}%
\caption{The radial part of the $\Lambda \Lambda$-component wave function $\Psi_{c=1}(\beta/2)$.}
\label{psi_beta_lambda}
\end{center}
\end{figure}

\begin{figure}[htbp]
\begin{center}
\subfigure[The wave function at $\beta =0.0001$ MeV$^{-1}$.]{%
\includegraphics[width=7.5cm, height=5.5cm]{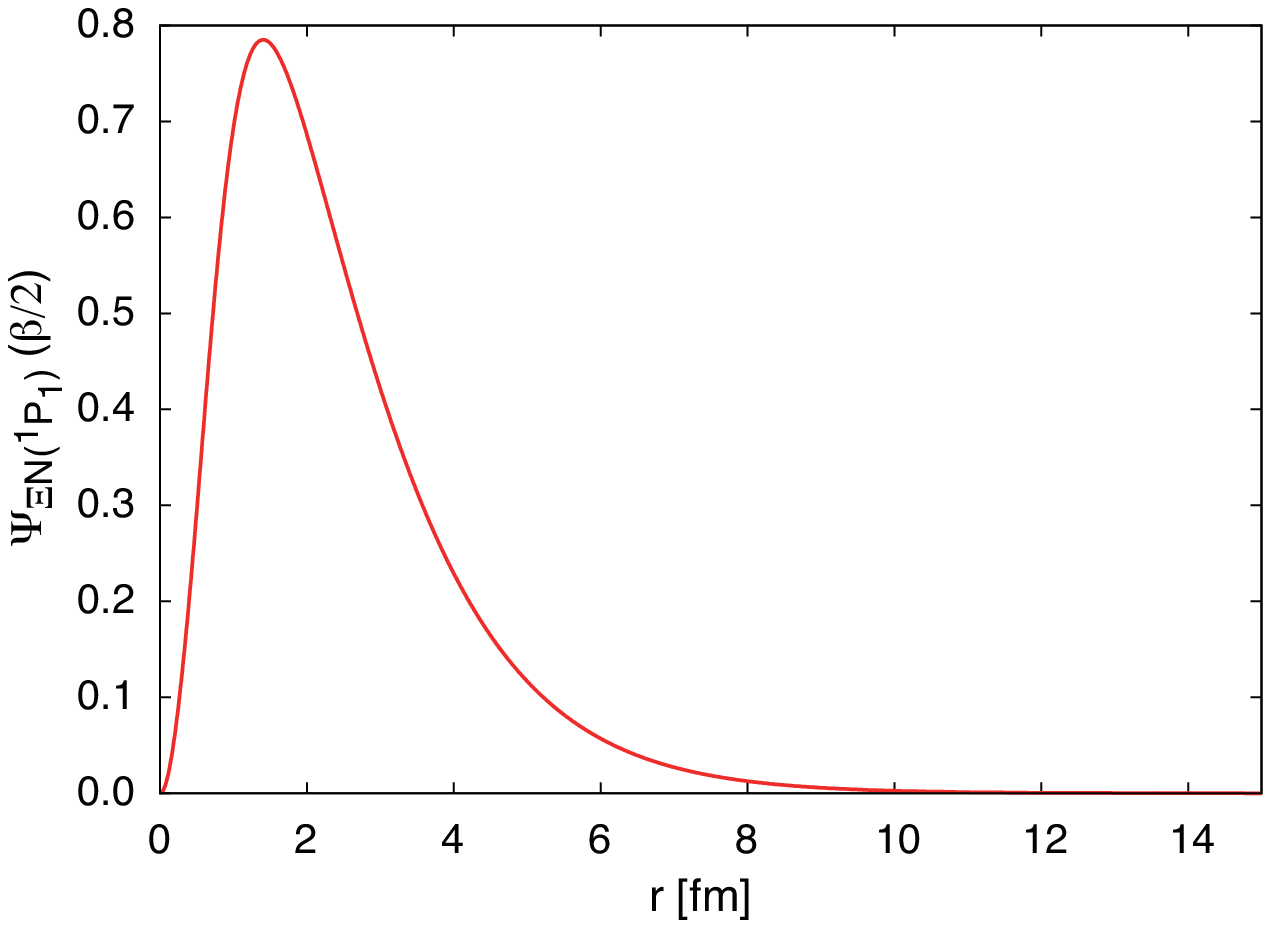}}%
\hspace{10mm}
\subfigure[The wave function at $\beta =0.001$ MeV$^{-1}$.]{%
\includegraphics[width=7.5cm, height=5.5cm]{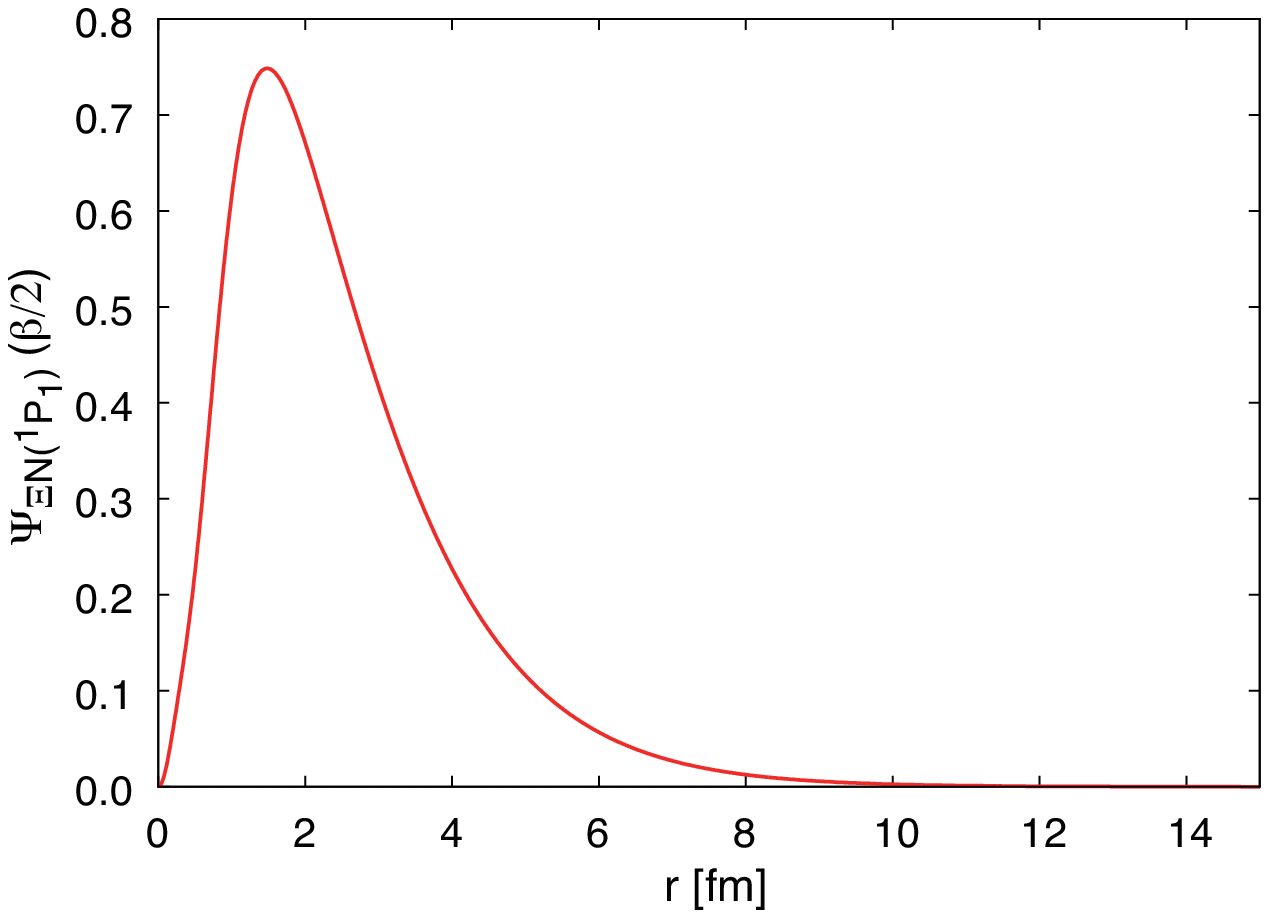}}%
\hspace{10mm}
\subfigure[The wave function at $\beta =0.01$ MeV$^{-1}$.]{%
\includegraphics[width=7.5cm, height=5.5cm]{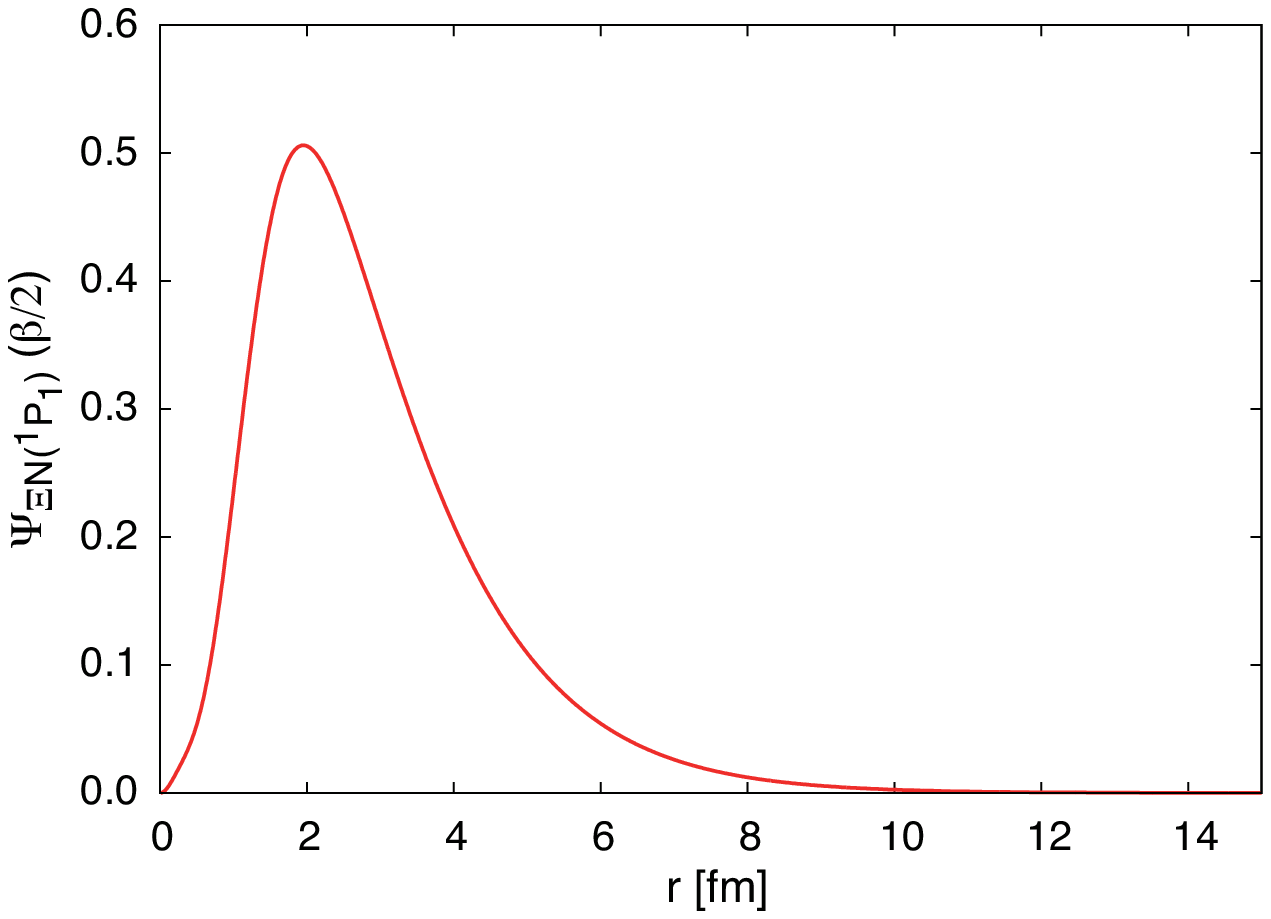}}%
\caption{The radial part of the $\Xi {\rm N}({}^1P_1)$-component wave function $\Psi_{c=2}(\beta/2)$.}
\label{psi_beta_NXi0}
\end{center}
\end{figure}

Figures \ref{psi_beta_lambda} and \ref{psi_beta_NXi0} show the behaviors of the imaginary-time evolution of the $\Lambda \Lambda (^3\textnormal{P}_1)$ and $\Xi $N$(^1\textnormal{P}_1)$ components of the $\vec{\Psi} (\beta /2)$ obtained by solving Eq.~(\ref{imaginary_eq}) with Eq.~(\ref{taylor}). 
In both figures, we find that the amplitudes of the wave functions are exuded toward the outer region as the imaginary-time evolution. While the amplitude of the $\Lambda \Lambda (^3\textnormal{P}_1)$ wave function $\Psi_2(\beta/2)$ becomes larger, as the imaginary-time evolution (see Fig.\ref{psi_beta_lambda}), the one of the $\Xi $N$(^1\textnormal{P}_1)$ wave function becomes smaller (see Fig.\ref{psi_beta_NXi0}).


\begin{figure}[htbp]
\begin{center}
\includegraphics[width=7.5cm, height=5.5cm]{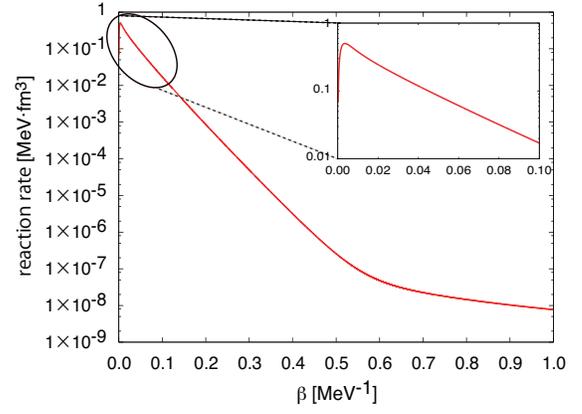}
\caption{The reaction rate of the H dibaryon as a function of the inverse temperature $\beta$.}
\label{reaction_rate}
\end{center}
\end{figure}

\begin{figure}[htbp]
\begin{center}
\subfigure[Calculated for $\beta \leq 10$ MeV$^{-1}$. ]{
\includegraphics[width=7.5cm, height=5.5cm]{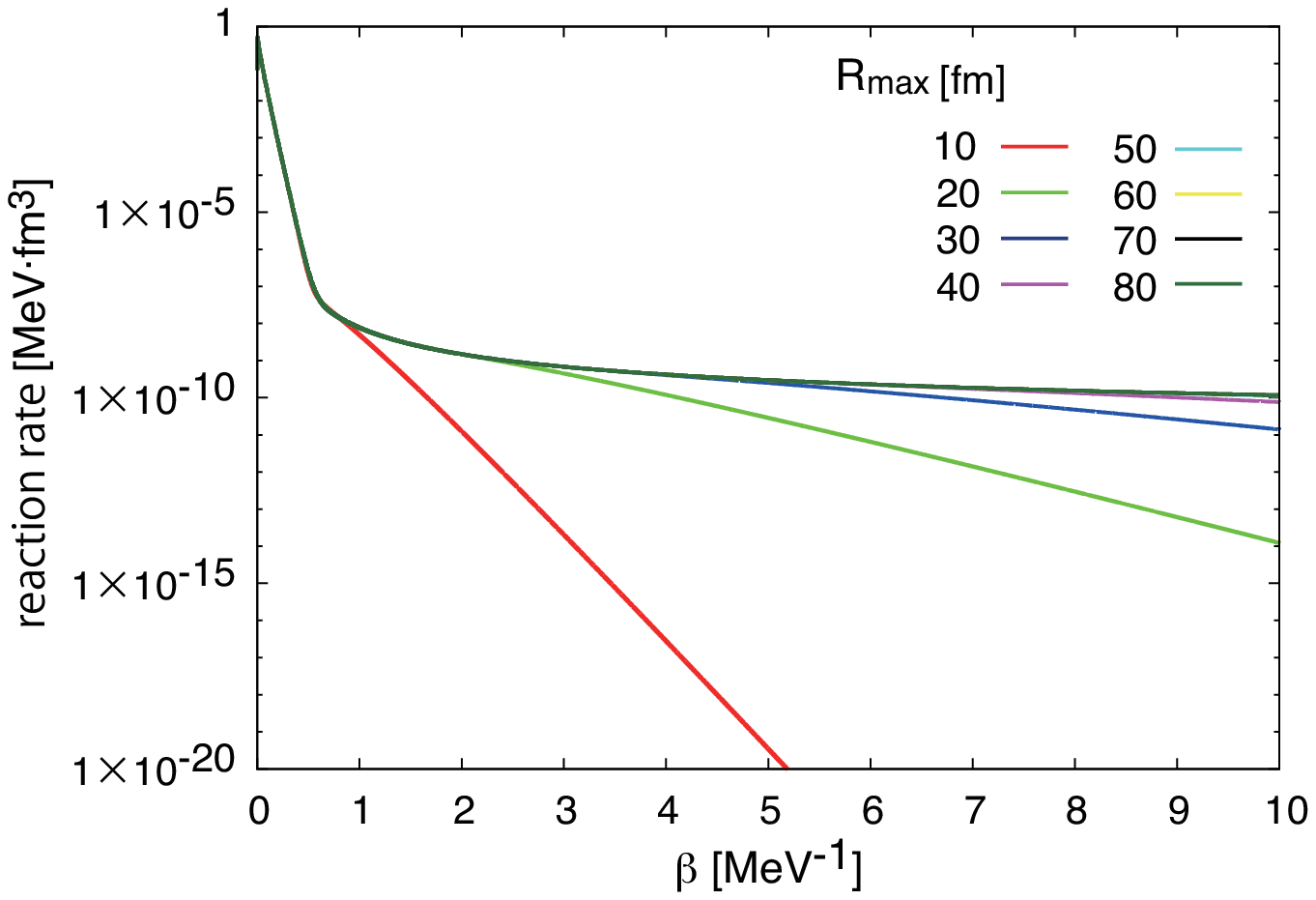}}
\hspace{10mm}
\subfigure[Enlarged view of (a) with $\beta \leq 1.0$ MeV$^{-1}$. ]{
\includegraphics[width=7.5cm, height=5.5cm]{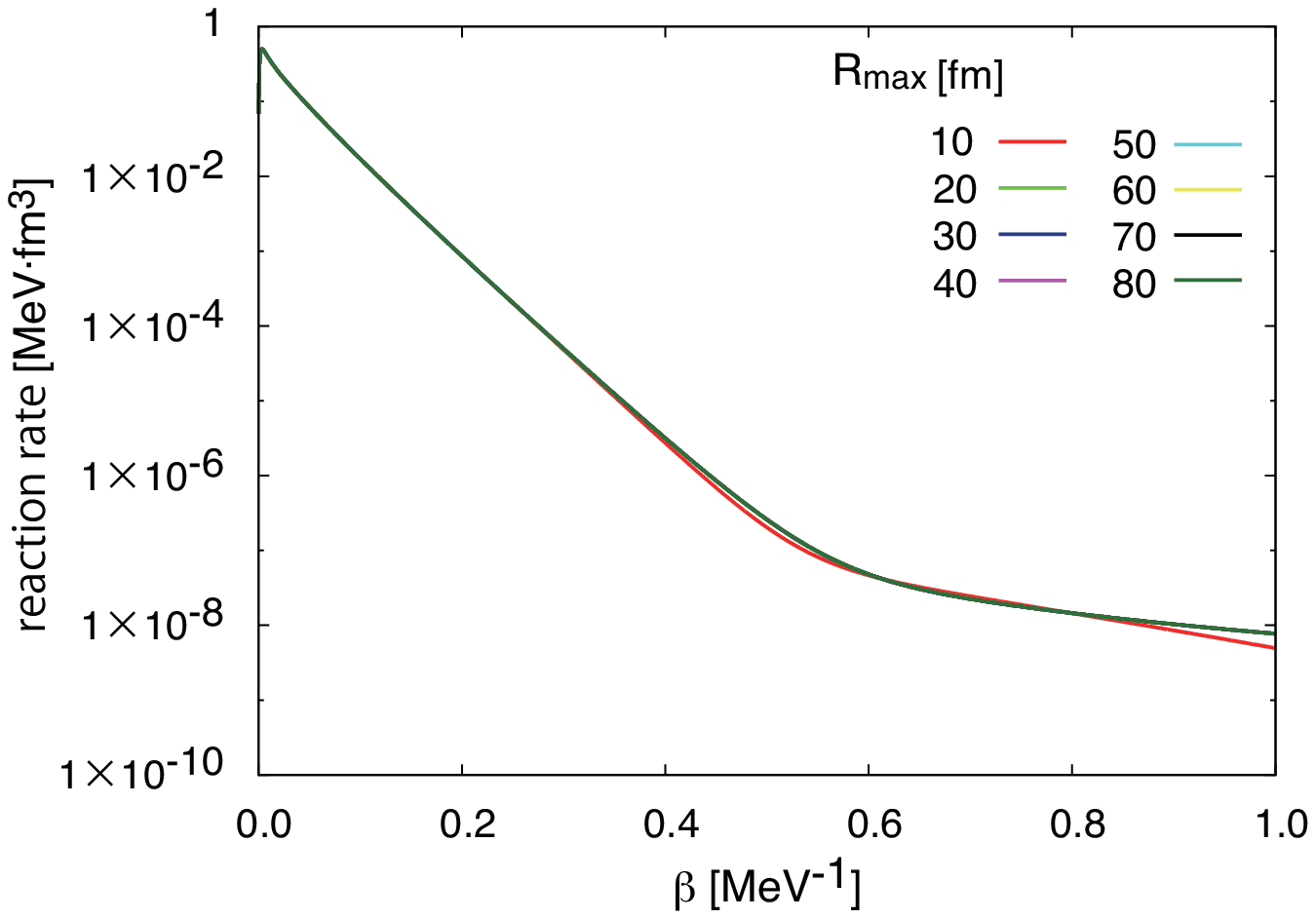}}
\caption{The convergence behavior of the calculated reaction rates with respect to the choice of the different radial box sizes $R_{\rm max}$.}
\label{reaction_rate_compare_radius}
\end{center}
\end{figure}

Figure~\ref{reaction_rate} is the calculated reaction rate in Eq.~(\ref{eq:itm2}) shown as a function of the inverse temperature $\beta$ [MeV], where the imaginary-time evolution of Eq.~(\ref{imaginary_eq}) and Eq.~(\ref{taylor}) is calculated in a large box size $R_{\rm max}=20$ fm. We check the convergence of the reaction rate with respect to the box size $R_{\rm max}$. In Fig.~\ref{reaction_rate_compare_radius}(a), we show the reaction rates calculated with the maximum size between $R_{\rm max}=10$ fm - 80 fm. We find that the reaction rate falls off too rapidly if the radial box size is not sufficiently large. This convergence behavior is the same as calculated in the previous articles with the use of the ITM~\cite{Yabana,Ya14}. As the imaginary time proceeds, i.e., the temperature decreases, the amplitude of the wave function gradually extends toward the outside region, since the lower energy states give significant contribution to the reaction rate at lower temperatures. We can see that the reaction rate is well converged up to $\beta=10$ [MeV$^{-1}$] if we take more than $R_{\rm max}=60$ fm. In Fig.\ref{reaction_rate_compare_radius}(b), the convergence behavior up to $\beta=1$ [MeV$^{-1}$] is shown, and we can see that adopting $R_{\rm max}=20$ fm gives sufficiently converged result. Since in this study we suppose heavy ion collision with thermal $\Lambda \Lambda$ gas with $\beta$ much less than 1 [MeV$^{-1}$], we adopt $R_{\rm max}=20$ fm in the subsequent calculations.

In Fig.~\ref{reaction_rate}, we find that the reaction rate of the H dibaryon is 0.3779\ [MeV$\cdot $fm$^3$] at $\beta =0.01$\ [MeV$^{-1}$] ($k_B T=100$\ [MeV]). Now we can estimate the number of the H dibaryon in the heavy ion collision from this reaction rate. Suppose that one heavy-ion collision produces 10 $\Lambda$ particles within the spatial size of $(5 [\textnormal{fm}])^3$ for the time of $10/c$ [s], with $c$ the light velocity. Then the created number of the H dibaryon is calculated to be $N=1.532 \times 10^{-3}$. This indicates that for approximately 650 collisions of heavy ions, one H-dybarion is produced.

\begin{figure}[htbp]
\begin{center}
\includegraphics[width=7.5cm, height=5.5cm]{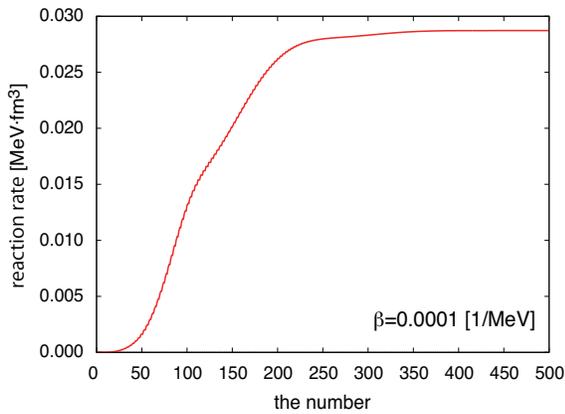}
\caption{The reaction rate at $\beta =0.0001$ MeV$^{-1}$ calculated with the ordinary method of Eq.~(\ref{ordinary}). The convergence behavior with respect to the number of initial states taken in the sum of Eq.~(\ref{ordinary}).}
\label{rate_sum}
\end{center}
\end{figure}

The ITM formula of Eq.~(\ref{eq:itm}) or Eq.~(\ref{eq:itm2}) could be obtained by transforming the ordinary form of the reaction rate Eq.~(\ref{ordinary}), where scattering wave functions should be handled explicitly as the initial states. The present formula is the extended version of the original ITM formula in Ref.~\cite{Yabana} to the case of coupled-channel calculations. Therefore it is important to check the correctness of the analytical formula of the coupled-channel ITM, by comparing the results calculated numerically according to both the fomulae Eq.~(\ref{eq:itm2}) and Eq.~(\ref{ordinary}). In order to calculate the reaction rate according to the ordinary method, we first derive the initial states $\psi_{ic}$ $(c=1,2,3)$ with the use of the conjugate gradient (CG) method in a spatial box size, which is here taken as 20\ [fm], with a radial grid size $\Delta r =0.01$ fm. The total number of the initial states is then $2000 \times 3$. Figure \ref{rate_sum} shows the convergence behavior of the reaction rate at $\beta =0.0001$\ [MeV$^{-1}$] for the adopted number of the initial states in the sum of the energy levels $i$ in Eq.~(\ref{ordinary}). We find that the reaction rate is converged when more than 400 initial states, i.e. $\psi_{i=1c} \sim \psi_{i=400c}$, are summed up in Eq.~(\ref{ordinary}). The eigenenergy of the 400th energy eigenstate is calculated to be $E_{i=400}=15748$ MeV.
%
\begin{figure}[htbp]
\begin{center}
\includegraphics[width=7.5cm, height=5.5cm]{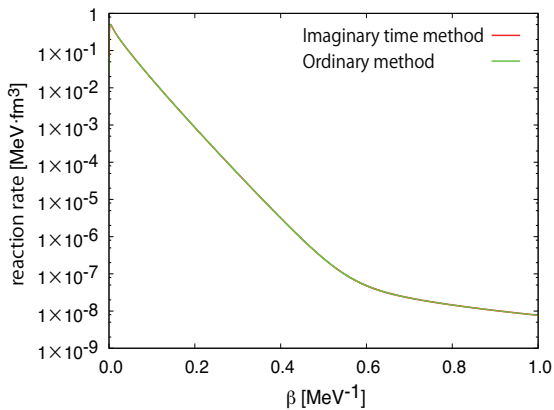}
\caption{The comparison of the reaction rates calculated with the ITM and the ordinary method.}
\label{reaction_rate_compare}
\end{center}
\end{figure}

In Fig.~\ref{reaction_rate_compare}, we show the reaction rate calculated as a function of $\beta$, in which the energy eigenstates up to $i_{\rm max}=400$ are summed up, in comparison with that obtained with the ITM. We can see that both results coincide with each other, indicating the correctness of the ITM formula derived within the coupled-channel framework. Here we mention an advantage of the ITM that it saves the computational time much more than the ordinary method. The former takes 57 seconds while the latter 608 seconds, at $\beta =1$ MeV$^{-1}$, i.e. 10 times faster than the ordinary method.

\begin{figure}[htbp]
\begin{center}
\includegraphics[width=7.5cm, height=5.5cm]{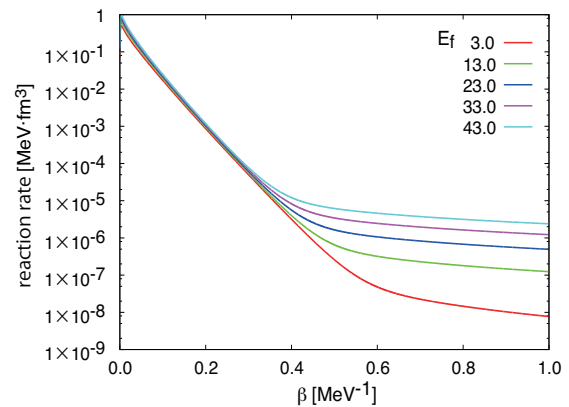}
\caption{Dependence of the reaction rate on the binding energy of the H dibaryon, which is varied by changing the depth parameter $V_0$ in Eq.~(\ref{eq:h}), with a fixed width-parameter value $r_{\rm H}=1.0$ fm.}
\label{reaction_rate_compare2}
\end{center}
\end{figure}

We next discuss the dependence of the reaction rate on the binding energy of the H dibaryon, which can be artificially tuned by changing the depth parameter of the potential $V_0$ in Eq.~(\ref{eq:h}) (see also Fig.~\ref{Hdibaryon_pot_para}). Figure \ref{reaction_rate_compare2} shows the reaction rate with different binding energies of the H dibaryon. We see that as the binding energy increases the reaction rate becomes large at lower temperature regions. However, at higher temperatures, $\beta \leq 0.3$ MeV$^{-1}$, which we are interested in in this study, the reaction rate is not sensitive to the variation of the binding energy of the H dibaryon. This implies that a similar result is expected for a resonance H dibaryon above $\Lambda \Lambda$ threshold, although in this calculation a bound state is assumed.
%
%

\begin{figure}[htbp]
\begin{center}
\includegraphics[width=7.5cm, height=5.5cm]{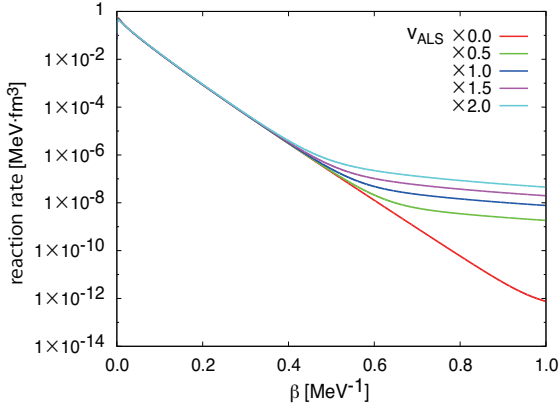}
\caption{Dependence of the reaction rate on the magnitudes of the ALS potential $v_{\textnormal{ALS}}$.}
\label{reaction_rate_ALS}
\end{center}
\end{figure}

We also investigate the dependence of the reaction rate on the choice of the ALS potential, by artificially varying the strength of the ALS potential $v_{\rm ALS}$ in Eq.~(\ref{eq:vals}). We show in Fig.~\ref{reaction_rate_ALS} the reaction rates calculated with the various strength values with which additional factors are multiplied. We find that at lower temperature region with $\beta \geq 0.4$ MeV$^{-1}$, the reaction rate is sensitive to the choice of the ALS potential, and the stronger ALS potential leads to the larger reaction rate, whereas at higher temperature region with $\beta \leq 0.4$ MeV$^{-1}$, the reaction rate is not sensitive to the choice of the ALS potential. This result is understood as follows: At low temperature region, the $\Lambda \Lambda$ states are dominant and the mixing of the $\Xi {\rm N}({}^1P_1)$ states with the $\Lambda \Lambda$ states is small. The direct transition of the $\Lambda \Lambda$ initial states into the H dibaryon is prohibited and only the ALS potential couples the $\Lambda \Lambda$ channel with the $\Xi {\rm N}({}^1P_1)$ channel that contributes to this transition, resulting in the strong sensitivity to the ALS potential. On the contrary, at higher temperatures, there are sufficient number of the $\Xi{\rm N}({}^1P_1)$ thermally mixed with $\Lambda \Lambda$ states. The mixing of the $\Xi {\rm N}$ states gives non-negligible contribution to the reaction rate, even without the ALS potential, leading to the insensitivity to the choice of the ALS potential. 

\begin{figure}[htbp]
\begin{center}
\includegraphics[width=7.5cm, height=5.5cm]{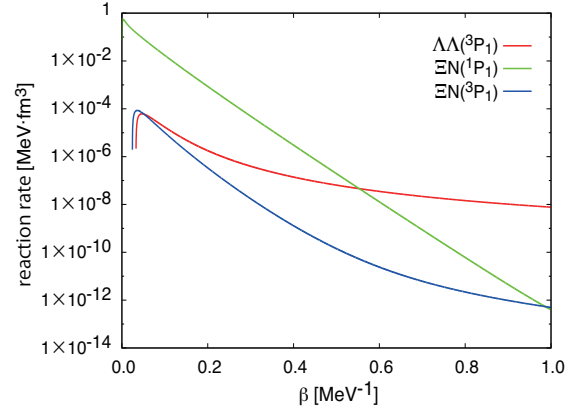}
\caption{Contributions of the reaction rate from the $\Lambda \Lambda$, $\Xi {\rm N}({}^1P_1)$, and $\Xi {\rm N}({}^3P_1)$ channels, $c_{\Lambda \Lambda ({}^3P_1)}$, $c_{\Xi {\rm N} ({}^1P_1)}$ and $c_{\Xi {\rm N} ({}^3P_1)}$, respectively, defined in Eq.~(\ref{eq:comp1}).}
\label{reaction_rate_contribution}
\end{center}
\end{figure}

Finally, we calculate the contributions to the reaction rate from the each component of the wave function $\vec{\Psi}(\beta)=(\Psi_1(\beta),\Psi_2(\beta),\Psi_3(\beta))$. In Eq.~(\ref{eq:itm2}), the reaction rate can be decomposed into the following three components, 
\begin{eqnarray}
c_{\Lambda \Lambda (^3\textnormal{P}_1)}&\propto & \sum_c \Braket{\Psi_1 \left( \frac{\beta }{2}\right) |\left( \frac{\hat{H}-E_f}{\hbar }\right)_{1c}^3 |\Psi_c \left( \frac{\beta }{2}\right) }, \nonumber \\
c_{\Xi \textnormal{N}(^1\textnormal{P}_1) }&\propto & \sum_c \Braket{\Psi_2 \left( \frac{\beta }{2}\right)  |\left( \frac{\hat{H}-E_f}{\hbar }\right)_{2c}^3 |\Psi_c \left( \frac{\beta }{2}\right) }, \nonumber \\
c_{\Xi \textnormal{N}(^3\textnormal{P}_1) }&\propto & \sum_c \Braket{\Psi_3 \left( \frac{\beta }{2}\right)  |\left( \frac{\hat{H}-E_f}{\hbar }\right)_{3c}^3 |\Psi_c \left( \frac{\beta }{2}\right) }, \nonumber \\ \label{eq:comp1}
\end{eqnarray}
where the following relation is satisfied,
\begin{eqnarray}
c_{\Lambda \Lambda (^3\textnormal{P}_1)}+c_{\Xi \textnormal{N}(^1\textnormal{P}_1) }+c_{\Xi \textnormal{N}(^3\textnormal{P}_1) }=r\left( \beta \right) .
\end{eqnarray}
We show in Fig.\ref{reaction_rate_contribution} these contributions from the three channels. We can see that at higher temperatures the contribution of the $\Xi $N$(^1\textnormal{P}_1)$ channel, $c_{\Xi \textnormal{N}({}^1P_1)}$ is the largest. This is because, according to the previous consideration, at higher temperatures a plenty of $\Xi {\rm N}({}^1P_1)$ states are mixed with the thermal $\Lambda \Lambda$ distribution, and the direct transition process from the $\Xi {\rm N}$ to the H dibaryon prevails over the process from $\Lambda \Lambda$ through the ALS coupling potential to the $\Xi {\rm N}({}^1P_1)$, and then to the H dibaryon. 
We also note that the wave-number factor $k$ of the emitted $\gamma$-ray in the transition operator for the $\Xi {\rm N}({}^1P_1)$ channel in Eq.~(\ref{eq:tra2}) makes $c_{\Xi \textnormal{N}(^1\textnormal{P}_1) }$ larger than $c_{\Xi \textnormal{N}(^3\textnormal{P}_1) }$.
On the contrary, in lower temperature region, there are few $\Xi {\rm N}$ states mixed with the $\Lambda \Lambda$ thermal distribution. The dominant contribution to the reaction rate is then from the $\Lambda \Lambda$ channel with a significant role of the ALS potential.

\begin{figure}[htbp]
\begin{center}
\includegraphics[width=7.5cm, height=5.5cm]{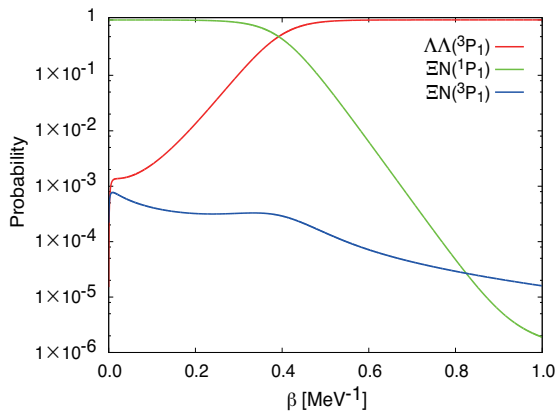}
\caption{Contributions of the probability from the $\Lambda \Lambda$, $\Xi {\rm N}({}^1P_1)$, and $\Xi {\rm N}({}^3P_1)$ channels, $p_{\Lambda \Lambda ({}^3P_1)}$, $p_{\Xi {\rm N} ({}^1P_1)}$ and $p_{\Xi {\rm N} ({}^3P_1)}$, respectively, defined in Eqs.~(\ref{eq:comp2-1})-(\ref{eq:comp2-3}).}
\label{reaction_rate_probability}
\end{center}
\end{figure}

We can understand this situation more clearly by the further analysis of the three-component wave function. In Fig.~\ref{reaction_rate_probability}, we shows the probabilities of the each component defined below,
\begin{eqnarray}
p_{\Lambda \Lambda (^3\textnormal{P}_1)}&=&\frac{\Braket{\Psi_1 \left( \frac{\beta }{2}\right) |\Psi_1 \left( \frac{\beta }{2}\right) }}{\sum_c \Braket{\Psi_c \left( \frac{\beta }{2}\right) |\Psi_c \left( \frac{\beta }{2}\right) }}, \label{eq:comp2-1} \\
p_{\Xi \textnormal{N}(^1\textnormal{P}_1) }&=&\frac{\Braket{\Psi_2 \left( \frac{\beta }{2}\right) |\Psi_2 \left( \frac{\beta }{2}\right) }}{\sum_c \Braket{\Psi_c \left( \frac{\beta }{2}\right) |\Psi_c \left( \frac{\beta }{2}\right) }}, \label{eq:comp2-2} \\
p_{\Xi \textnormal{N}(^3\textnormal{P}_1) }&=&\frac{\Braket{\Psi_3 \left( \frac{\beta }{2}\right) |\Psi_3 \left( \frac{\beta }{2}\right) }}{\sum_c \Braket{\Psi_c \left( \frac{\beta }{2}\right) |\Psi_c \left( \frac{\beta }{2}\right) }}, \label{eq:comp2-3}
\end{eqnarray}
where the following relation is satisfied,
\begin{eqnarray}
p_{\Lambda \Lambda (^3\textnormal{P}_1)}+p_{\Xi \textnormal{N}(^1\textnormal{P}_1) }+p_{\Xi \textnormal{N}(^3\textnormal{P}_1) }=1.
\end{eqnarray}
We can see that it is found that, as is consistent with the previous figure, at high temperatures the $\Xi {\rm N}({}^1P_1)$ component is dominant, and at the lower temperatures the $\Lambda \Lambda$ component becomes the largest. As the decrease of temperature, the probability of the $\Lambda \Lambda$ component rapidly gets higher, and over $\beta \approx 0.4$ MeV$^{-1}$ it becomes the largest, whereas below this temperature the $\Xi {\rm N}({}^1P_1)$ component rapidly disappears.
\section{Summary and Conclusion}
In summary, we have applied the imaginary time method to calculate the radiative fusion rates of $\Lambda\Lambda$ into the H dibaryon at finite temperature.
Mixings of the $\Xi $N$(^3\textnormal{P}_1)$ and $\Xi $N$(^1\textnormal{P}_1)$ channels with the $\Lambda\Lambda (^3\textnormal{P}_1)$ channel are considered in the initial thermal state.
The E1 transition to the $\Xi $N$(^1\textnormal{S}_0)$ bound state that represents the H dibaryon is calculated.
The imaginary time method is applied so that the sum over all the excited states in the thermal initial state can be taken into account
without computing the scattering wave functions.

A representative calculation is done for a bound H dibaryon with the binding energy 3 MeV below the $\Lambda\Lambda$ threshold for the initial temperature of 100 MeV, giving the transition rate, 0.38\ [MeV$\cdot $fm$^3$]. 
This corresponds to a rate that may produce one H dibaryon in a few hundred HI collisions through the E1 transition, assuming a collision
produces 10 $\Lambda$'s in a few fm volume region of temperature about 100 MeV.
It is found that the effect of the channel mixing is significant especially at high temperature,
where the difference of the threshold energies between $\Lambda\Lambda$ and $\Xi$N is not important. 
On the other hand, at a lower temperature, the transition rates are sensitive to the mixing potential.
It should be noted that the E1 transition is dominated from the $\Xi $N$(^1\textnormal{P}_1)$ channel, 
which mixes with the $\Lambda\Lambda (^3\textnormal{P}_1)$ state only through the antisymmetric spin-orbit (ALS) potential.
Thus the transition rates at low temperatures are sensitive to the strength of the ALS potential.
We have found that the transition rate is larger for the deeper bound state at low temperatures.
In contrast,  the reaction rates are not sensitive to the choices of parameters when the temperature is above around 50 MeV.
Although our present calculation is applied only to the bound H state,  
we expect a similar production rate, even if the H dibaryon is a resonance above the $\Lambda \Lambda$ threshold.

Our calculation is compared to the standard method, which explicitly sums the initial scattering states, and
it is found that the results are consistent with each other.
On the other hand, the computational time in the imaginary time method is much shorter (about one order) than the ordinary method.
Thus, we have confirmed that the imaginary time method for the multi-channel hadronic transitions can be well performed.
\begin{acknowledgments}
This work was supported by JSPS KAKENHI Grant Numbers  23224006, 25400288, 25247036, RIKEN iTHES Project, and HPCI Strategic Program of Japanese MEXT.
\end{acknowledgments}

\appendix*
\section{Parameter sets of YY and YN interaction}

We show in Table \ref{table_1} the parameter sets of the $G$-matrix-type YY and YN interaction, which are expressed in the Gaussian form in Eq.~(\ref{gauss}). For the central, SLS and tensor forces, we adopt the Nijmegen ESC08 potential, while for the ALS force the Nijmegen NSC97-f is chosen. The width and strength parameters of the Gaussian form are shown for each $v_{cc'}$ channel with $c,c'=1,2,3$. As described in the text, the ALS force appears only in $v_{12}$ and $v_{23}$, for which no terms contribute other than the ALS term.

\begin{table*}[htbp]
\caption{The width and strength parameters of the YY and YN potential written in terms of the Gaussian form in Eq.~(\ref{gauss}) are shown. The parameter sets of ESC08c~\cite{Ya15} and NSC97-f~\cite{hiyamasan_ronbun} are adopted for the central, SLS and tensor forces, and the ALS force, respectively.}\label{table_1}
\begin{tabular}{cccccccccc}
\hline\hline
\raisebox{-1.8ex}[0pt][0pt]{$v_{ij}$} & \raisebox{-1.8ex}[0pt][0pt]{$k$} & \raisebox{-1.8ex}[0pt][0pt]{$r_k^{(C)}$ [fm]} & \raisebox{-1.8ex}[0pt][0pt]{$v_k^{(C)}$ [MeV]} & \raisebox{-1.8ex}[0pt][0pt]{$r_k^{(S)}$ [fm]} & \raisebox{-1.8ex}[0pt][0pt]{$v_k^{(S)}$ [MeV]} & \raisebox{-1.8ex}[0pt][0pt]{$r_k^{(T)}$ [fm]} & \raisebox{-1.8ex}[0pt][0pt]{$v_k^{(T)}$ [MeV]} & \raisebox{-1.8ex}[0pt][0pt]{$r_k^{(A)}$ [fm]} & \raisebox{-1.8ex}[0pt][0pt]{$v_k^{(A)}$ [MeV]} \\
 &  &  &  &  &  &  &  &  &  \\
\hline
\multicolumn{1}{l}{$\Lambda \Lambda({}^3P_1)$-$\Lambda \Lambda({}^3P_1)$} & 1 & $0.2274$ & $-1.707\times 10^3$ & 0.05465 & $-2.297\times 10^{3}$ & 0.07595 & $-1.018\times 10^{2}$ &  &  \\
 & 2 & $0.3213$ & $\ \ 1.100\times 10^{4}$ &  0.09786 & $-1.005\times 10^{3}$ & 0.1512 & $\ \ 2.113\times 10^{2}$ &  &  \\
 & 3 & $0.4538$ & $-4.355\times 10^{3}$ & 0.1752 & $-4.807\times 10^{2}$ &  0.3011 & $-2.911\times 10^{2}$ &  &  \\
\raisebox{-1.8ex}[0pt][0pt]{$v_{11}$} & 4 & $0.6411$ & $\ \ 1.028\times 10^{3}$ & 0.3137 & $\ \ 2.178\times 10^{3}$ &  0.5997 & $\ \ 1.520\times 10^{2} $ &  &  \\
 & 5 & $0.9057$ & $-8.899\times 10^{1}$ & 0.5618 & $-1.913\times 10^{2}$  & 1.194 & $-4.329\times 10^{0}$ &  &  \\
 & 6 &  &  & 1.005 & $\ \ 1.239\times 10^{1}$ &  &  &  &  \\
 & 7 &  &  & 1.801 & $-7.992\times 10^{-1}$ &  &  &  &  \\
 & 8 &  &  & 3.225 & $\ \ \ 7.613\times 10^{-2}$ &  &  &  &  \\
\hline
\multicolumn{1}{l}{$\Xi$N$(^1P_1)$-$\Xi$N$(^1P_1)$} & 1 & 0.1673 & $\ \ 1.180\times 10^{2}$ &  &  &  &  &  &  \\
 & 2 & 0.2778 & $-3.910\times 10^{3}$ &  &  &  &  &  &  \\
$v_{22}$ & 3 & 0.4613 & $\ \ 1.021\times 10^{3}$ &  &  &  &  &  &  \\
 & 4 & 0.7659 & $\ \ 5.406\times 10^{2}$ &  &  &  &  &  &  \\
 & 5 & 1.271 & $-3.311\times 10^{1}$ &  &  &  &  &  &  \\
\hline
\multicolumn{1}{l}{$\Xi$N$({}^3P_1)$-$\Xi$N$({}^3P_1)$} & 1 & 0.1817 & $-1.449\times 10^{3} $ & 0.1141 & $-2.052\times 10^{3}$ &  0.1456 & $\ \ 2.648\times 10^{1}$ &  &  \\
 & 2 & 0.3165 & $\ \ 3.956\times 10^{3}$ & 0.2696 & $-3.974\times 10^{3}$ & 0.2789 & $-5.224\times 10^{2}$ &  &  \\
$v_{33}$ & 3 &  0.5512 & $-3.088\times 10^{1}$ & 0.6366 & $\ \ 1.621\times 10^{2}$ & 0.5341 & $\ \ 4.469\times 10^{2}$ &  &  \\
 & 4 &  0.9599 & $\ \ 1.982\times 10^{1}$ & 1.503 & $\ \ 8.211\times 10^{0}$ & 1.022 & $\ \ 4.907\times 10^{1}$ &  &  \\
 & 5 & 1.671 & $-1.217\times 10^{1}$ & 3.550 & $-4.802\times 10^{-1}$ & 1.958 & $\ \ 3.771\times 10^{0}$ &  &  \\
\hline
\multicolumn{1}{l}{$\Lambda \Lambda({}^3P_1)$-$\Xi$N$({}^3P_1)$} & 1 & 0.06457 & $-9.471\times 10^{2} $ & 0.05748 & $3.213\times 10^{2}$ & 0.2276 & $-4.925\times 10^{2}$ &  &  \\
 & 2 & 0.1308 & $\ \ 1.658\times 10^{3}$ & 0.1272 & $\ \ 6.710\times 10^{1}$ & 0.3308 & $\ \ 9.332\times 10^{2}$ &  &  \\
$v_{13}(=v_{31})$ & 3 & 0.2653 & $-4.896\times 10^{3}$ & 0.2817 & $-7.415\times 10^{3}$ &  0.4806 & $-5.395\times 10^{2}$ &  &  \\
 & 4 & 0.5378 & $\ \ 8.913\times 10^{2}$ & 0.6236 & $\ \ 3.363\times 10^{2}$ & 0.6983 & $\ \ 9.507\times 10^{1}$ &  &  \\
 & 5 &  1.090 & $-8.803\times 10^{1}$ & 1.380 & $\ \ 3.634\times 10^{0}$ & 1.014 & $\ -9.035\times 10^{-1}$ &  &  \\
\hline
\multicolumn{1}{l}{$\Lambda \Lambda({}^3P_1)$-$\Xi$N$({}^1P_1)$} & 1 &  &  &  &  &  &  & $0.75$ & $-64.50$ \\
$v_{12}(=v_{21}=v_{23}=v_{32})$ & 2 &  &  &  &  &  &  & $0.40$ & $-686.6$ \\
\hline
\end{tabular}
\end{table*}




\begin{thebibliography}{99}
%
\bibitem{Jaffe}
 R.\;L.\;Jaffe,\;Phys.\;Rev.\;Lett.\;\textbf{38},\;195\;(1977).
\bibitem{Oka:1983ku} 
  M.~Oka, K.~Shimizu and K.~Yazaki,
  Phys.\ Lett.\ B {\bf 130}, 365 (1983).
%
\bibitem{Takeuchi:1991}
S.~Takeuchi and M.~Oka, Phys.\ Rev.\ Lett.\; \textbf{66},\;1271\;(1991).
\bibitem{Sakai:1999qm} 
  T.~Sakai, K.~Shimizu and K.~Yazaki,
  Prog.\ Theor.\ Phys.\ Suppl.\  {\bf 137}, 121 (2000)
  [nucl-th/9912063].
\bibitem{NPLQCD}
S.\;R.\;Beane,\;\textit{et al.},\;Phys.\;Rev.\;Lett.\;\textbf{106},\;162001\;(2011).
%
\bibitem{HALQCD}
T.\;Inoue,\;\textit{et al.},\;Phys.\;Rev.\;Lett.\;\textbf{106},\;162002\;(2011).
%
\bibitem{Takahashi} 
  H.~Takahashi, {\it et al.},
  Phys.\ Rev.\ Lett.\  {\bf 87}, 212502 (2001).
\bibitem{Yoon}
C.~J.~Yoon, {\it et al.} (KEK-E522 collab.), Phys.\;Rev.\; \textbf{C 75},\;022201\;(2007).
\bibitem{EXHIC} 
  S.~Cho {\it et al.}  [ExHIC Collaboration],
  Phys.\ Rev.\ Lett.\  {\bf 106}, 212001 (2011)
%
\bibitem{Yabana}
K.\;Yabana\;and\;Y.\;Funaki,\;Phys.\;Rev.\;C.\;\textbf{85},\;055803\;(2012).
%
\bibitem{Ya14}
T. Akahori, Y. Funaki and K. Yabana, arXiv: 1401.4390.
\bibitem{Sa52}
E. E. Salpeter, Astrophys. J. {\bf 115}, 326 (1952).
\bibitem{Ho54}
F. Hoyle, Astrophys. J. Suppl. {\bf 1}, 121 (1954).
\bibitem{Co57}
C. W. Cook {\it et al.}, Phys. Rev. {\bf 107}, 508 (1957).
\bibitem{hiyamasan_ronbun}
T. A. Rijken, V. G. J. Stoks, and Y. Yamamoto, Phys. Rev. C {\bf 59}, 21 (1999);
E.\; Hiyama,\;\textit{et al.},\;Phys.\;Rev\;Lett.\;\textbf{85},\;270\;(2000); Y. Yamamoto, T. Motoba and T. A. Rijken, Prog. Theor. Phys. Suppl., {\bf 185}, 72 (2010).

\bibitem{Ya15}
Y. Yamamoto {\it et al.}, private communications.
%
%
%
\end{thebibliography}
\end{document}